\documentclass[fleqn,twoside]{article}
\usepackage{espcrc2}

\usepackage[latin1]{inputenc}
\usepackage[tbtags]{amsmath}
\usepackage{amssymb}

\usepackage{graphicx}

\newcommand{\Pe}{\mathrm e}
\newcommand{\Pep}{\mathrm {e^+}}
\newcommand{\Pem}{\mathrm {e^-}}

\newcommand{\Pt}{\mathrm t}

\newcommand{\PW}{\mathrm W}
\newcommand{\PZ}{\mathrm Z}
\newcommand{\PH}{\mathrm H}
\newcommand{\MW}{M_\PW}
\newcommand{\MZ}{M_\PZ}
\newcommand{\MH}{M_\PH}
\newcommand{\Mt}{m_\Pt}
\newcommand{\Oa}{{{\cal{O}}(\alpha)}}
\newcommand{\Oas}{{{\cal{O}}(\alpha_s)}}
\newcommand{\gtt}{\ensuremath{g_{\Pt\bar\Pt\PH}}}
\newcommand{\fb}{\unskip\,\mathrm{fb}}

\def\reffi#1{Figure~\ref{#1}}
\def\citere#1{Ref.~\cite{#1}}
\def\citeres#1{Refs.~\cite{#1}}

\newcommand{\eennh}{\ensuremath{\Pep\Pem\to\nu\bar\nu\PH}}
\newcommand{\eetth}{\ensuremath{\Pep\Pem\to\Pt\bar\Pt\PH}}
\newcommand{\eeffh}{\ensuremath{\Pep\Pem\to \mathrm{f}\Bar{\mathrm{f}} \PH}}
\newcommand{\ee}{\ensuremath{\Pep\Pem}}

\newcommand{\GeV}{\ensuremath{\,\text{GeV}}}
\newcommand{\TeV}{\ensuremath{\,\text{TeV}}}
\newcommand{\Gmu}{\ensuremath{G_\mu}}

\title{Electroweak corrections to \eeffh\hspace{1mm}\thanks{This work was
    supported in part by the Swiss Bundesamt für Bildung und
    Wissenschaft and by the European Union under contract
    HPRN-CT-2000-00149.}}

\author{A. Denner\address{Paul Scherrer Institut, Würenlingen und Villigen,
        CH-5232 Villigen PSI, Switzerland},
        S. Dittmaier\address[MPI]{Max-Planck-Institut für Physik 
          (Werner-Heisenberg-Institut),
          D-80805 München, Germany},
        M. Roth\addressmark[MPI]
        and M. M. Weber\address{Dipartimento di Fisica Teorica, 
          Università di Torino,
          Via Giuria 1, 10125 Torino, Italy}}

\begin{document}

\begin{abstract}
  Some of the most interesting Higgs-production processes at future
  $\ee$ colliders are of the type $\eeffh$. We present a calculation
  of the complete $\Oa$ corrections to these processes in the Standard
  Model for final-state neutrinos and top quarks. Initial-state
  radiation beyond $\Oa$ at the leading-logarithmic level as well as
  QCD corrections are also included.  The electroweak corrections turn
  out to be sizable and reach the order of $\pm 10 \%$ and will thus
  be an important part of precise theoretical predictions for future
  $\ee$ colliders.
\end{abstract}

\maketitle

\section{Introduction}

One of the main future tasks in particle physics will be the
investigation of the mechanism of electroweak symmetry breaking in
general and the discovery of the Higgs boson and the determination of its
properties in particular. Since the Higgs-boson mass is expected to be
in the range from the lower experimental bound of $114.4 \GeV$ up to $1
\TeV$, with a light Higgs mass (below $\sim 200\GeV$) favoured by
electroweak precision data, the LHC will be able to discover it in the
full mass range, provided it exists and has no exotic properties.
However, for the complete determination of its profile, including its
couplings to fermions and gauge bosons, experiments in the clean
environment of an $\ee$ linear collider are indispensable.

Here we concentrate on the associated production of a Higgs boson
together with a pair of neutrinos or top quarks in $\ee$
annihilation, which are among the most interesting Higgs-boson
production processes at future $\ee$ linear colliders.

\section{The process \boldmath{$\Pep\Pem \to \nu \bar{\nu} \PH$}}

At $\ee$ colliders the two main Higgs production processes are the
Higgs-strahlung and W-boson-fusion processes.  In the Higgs-strahlung
process the Higgs boson is radiated off a $\PZ$ boson, with the
corresponding cross section rising sharply at the threshold, located at
a centre-of-mass (CM) energy of $\sqrt{s} = \MZ + \MH$, to a maximum a
few tens of $\GeV$ above the threshold energy and then falling off as
$1/s$.  In the W-boson-fusion process the Higgs boson is produced via
fusion of two W~bosons, each emitted from an incoming electron/positron.
The corresponding cross section grows as $\ln s$ and thus is the
dominant production mechanism at large energies.  Both production
mechanisms appear in the process $\Pep \Pem \to \nu_l \Bar{\nu}_l
\PH$, with $l = \Pe,\mu$, or $\tau$, though the W-boson-fusion process
is only present for $l = \Pe$.

\begin{figure*}[t]
\centerline{
\includegraphics[bb=80 440 290 626, width=.46\textwidth]{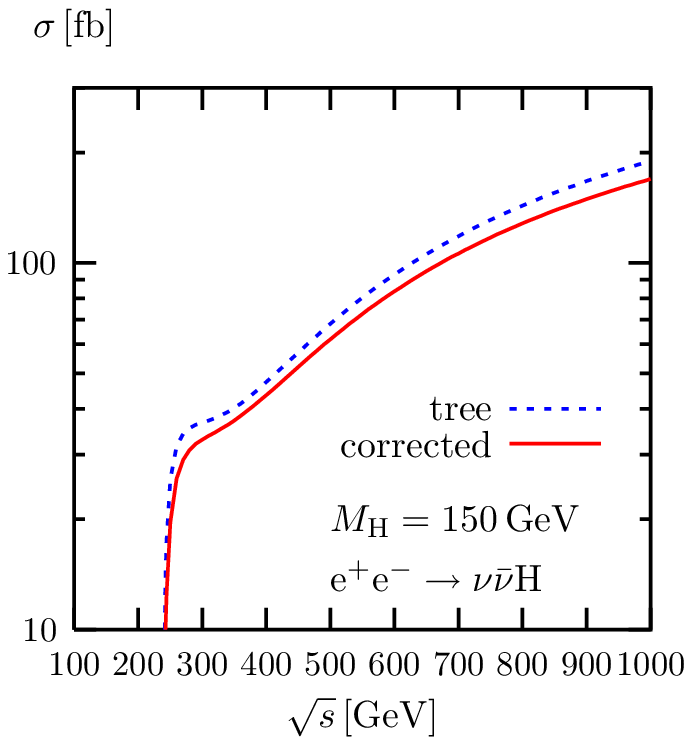}
\qquad
\includegraphics[bb=80 440 290 626, width=.46\textwidth]{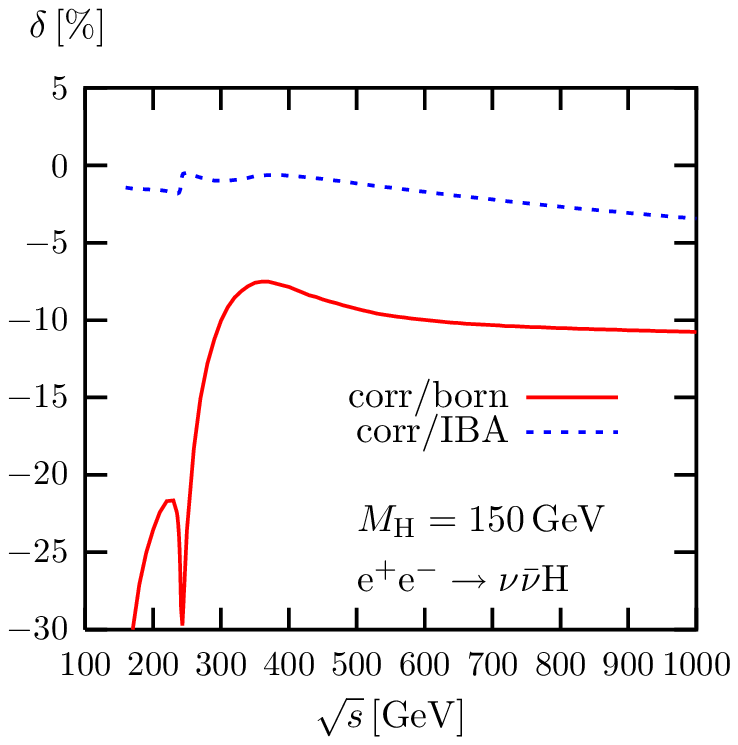}
}
\caption{Lowest-order and corrected cross sections (l.h.s.) as well as 
  relative corrections with respect to Born result and improved Born
  approximation (r.h.s.) in the $\Gmu$ scheme for a Higgs-boson mass
  $\MH=150\GeV$}
\label{fig:nnh}
\end{figure*}

For the process $\Pep\Pem\to\PZ\PH$ the $\Oa$ electroweak radiative
corrections have been calculated many years ago in
\citere{Fleischer:1982af}. Furthermore a Monte Carlo algorithm for the
calculation of the real photonic corrections to this process was
described in \citere{Berends:dw}. 
For the full process $\eennh$ there has been a lot of activity
regarding the electroweak corrections recently.  Within the Minimal
Supersymmetric Standard Model (MSSM) the fermion and sfermion loop
contributions have been evaluated in
\citeres{Eberl:2002xd,Hahn:2002gm}.  Analytical results for the
one-loop corrections in the SM have been obtained in
\citere{Jegerlehner:2002es}, though no numerical results have been
given there. Finally, calculations of the complete $\Oa$ electroweak
corrections to $\eennh$ in the SM have been performed in
\citeres{Denner:2003yg,Belanger:2002ik}. A comparison of these
calculations has revealed an agreement within $0.3\%$, which is of the
same order as the integration error of \citere{Belanger:2002ik}.
Very recently also results on corrections to the Z-boson-fusion process
$\Pep\Pem \to \Pep\Pem\PH$ have been presented in \citere{Boudjema:2004ba}.

A sketch of the calculational setup and a summary of the main results
of our calculation \cite{Denner:2003yg} of the complete electroweak
corrections is given in the following.  Apart from the $\Oa$
corrections we have also included the leading-logarithmic part of the
higher-order initial-state radiation (ISR) using the
structure-function approach.  Furthermore by using the $\Gmu$ scheme
we have absorbed corrections proportional to $\Mt^2/\MW^2$ in the
fermion--W-boson couplings and the running of $\alpha(Q^2)$ from
$Q^2=0$ to the electroweak scale.  The calculation has been performed
mostly using standard techniques. However, the appearance of pentagon
diagrams potentially leads to numerical instabilities related to
leading inverse Gram determinants. We have therefore used the
reduction scheme of \citere{Denner:2002ii}.  For the extraction of the
soft and collinear singularities in the real corrections we have used
both the dipole subtraction method \cite{Dittmaier:1999mb,Roth:1999kk}
and phase-space slicing following closely \citere{Bohm:1993qx}.  Two
independent calculations have been made resulting in two independent
computer codes for the numerical evaluation, one employing a
multi-channel Monte-Carlo generator similar to
\citeres{Roth:1999kk,Denner:1999gp,Dittmaier:2002ap} for the
phase-space integration, the other one using {\sc Vegas}
\cite{Lepage:1977sw}.

\begin{figure*}[t]
\centerline{
\includegraphics[bb=80 440 290 626, width=.46\textwidth]{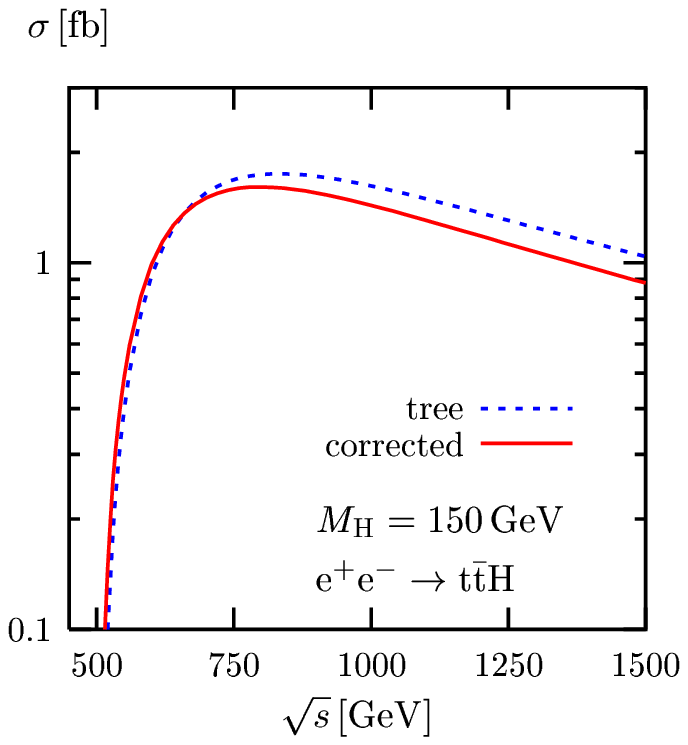}
\qquad
\includegraphics[bb=80 440 290 626, width=.46\textwidth]{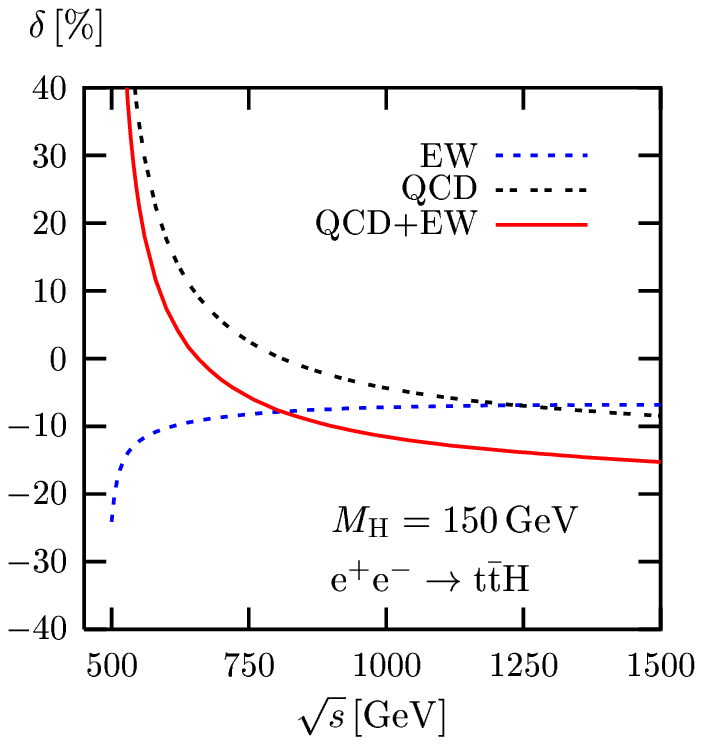}
}
\caption{Lowest-order and corrected cross sections (l.h.s.) as well as
  relative corrections (r.h.s.) in the $\Gmu$ scheme for a Higgs-boson
  mass $\MH=150\GeV$}
\label{fig:tth}
\end{figure*}

The results for the total cross section in lowest order and including
the radiative corrections are shown in \reffi{fig:nnh} on the l.h.s.\ 
as a function of the CM energy for $\MH = 150\GeV$.  The relative
corrections shown on the r.h.s are large ($\lesssim -20\%$) and vary
stronlgy in the ZH-threshold region while they are flat and about
$-10\%$ for energies above $500\GeV$. They are always negative because
they are dominated by initial-state radiation and the cross section is
monotonously rising.  We have also constructed an improved Born
approximation (IBA) which incorporates the leading-logarithmic part of
the ISR using structure functions and furthermore contains the leading
$\Mt^2 / \MW^2$ corrections from the WWH-vertex.  As shown in
\reffi{fig:nnh} (r.h.s), the residual relative corrections normalized
to the IBA are about 1--3\%. Although they are systematically smaller
than the corrections relative to the lowest order in the $\Gmu$ scheme,
the inclusion of the full $\Oa$ corrections is necessary for a
precision analysis.

\section{The process \boldmath{$\Pep\Pem \to \Pt\bar\Pt\PH$}}

We have also investigated the process $\eetth$, which is interesting
since it permits a direct access to the top-quark Yukawa coupling
$\gtt$, which is by far the largest Yukawa coupling ($\gtt \approx
0.5$) in the SM.  This is possible because the process proceeds
mainly through Higgs-boson emission off top quarks, while emission
from intermediate Z bosons plays only a minor role if the Higgs-boson
mass is not too large, i.e.\ $\MH\sim 100$--$200\GeV$.  For a light
Higgs boson with a mass around $\MH\sim 120\GeV$, a precision of about
$5\%$ can be reached at an $\Pep\Pem$ linear collider operating at
$\sqrt{s} = 800\GeV$ with a luminosity of $\int L\,\mathrm{d} t \sim
1000\fb^{-1}$ \cite{Baer:1999ge}.  An even better accuracy can be
obtained by combining the $\Pt\bar{\Pt}\PH$ channel with information
from other Higgs-production and decay processes in a combined fit
\cite{Battaglia:2000jb}.

Within the SM the $\Oas$ corrections have been calculated for the
dominant photon-exchange channel in \citere{Dawson:1998ej}, while the
full set of diagrams has been evaluated in \citere{Dittmaier:1998dz}.
The $\Oas$ corrections to the photon-exchange channel in the MSSM have
been considered in \citere{Dawson:1998qq}. In
\citere{Dittmaier:2000tc} all QCD diagrams have been taken into
account, while the SUSY-QCD corrections have been worked out in
\citere{Zhu:2002iy}.  The evaluation of the electroweak $\Oa$
corrections in the SM has made considerable progress recently. Results
have been presented in
\citeres{You:2003zq,Belanger:2003nm,Denner:2003ri}, with agreement
between \citeres{Belanger:2003nm,Denner:2003ri} while
\citere{You:2003zq} shows deviations close to threshold and at high
energies.

We finally present some results of our calculation of the $\Oa$
electroweak and the $\Oas$ QCD corrections. Though the calculation of
the virtual corrections for this process is much more involved than
for the process $\eennh$, the same calculational techniques could be
used.

Results for the total cross section in lowest order and the corrected
cross section including both the electroweak and QCD corrections are
shown in \reffi{fig:tth} on the l.h.s.  Away from the kinematic
threshold at $\sqrt{s}=2 \Mt+\MH$ the size of the cross section is
typically a few $\fb$, with a maximum at about $800\GeV$.  On the
r.h.s.\ of \reffi{fig:tth} the relative corrections are shown.  The
QCD corrections are large and positive close to threshold where
soft-gluon exchange in the $\Pt\bar\Pt$ system leads to a Coulomb-like
singularity.  For larger energies the QCD corrections decrease,
eventually turn negative and reach about $-8\%$ at an energy of
$\sqrt{s} = 1.5\TeV$.  The electroweak corrections are about $-10\%$
and only vary weakly with energy away from the threshold region, and
are thus of a comparable size as the QCD corrections. Close to
threshold they reach about $-20\%$ due to the large ISR QED
corrections in this region.  The behaviour of the combined electroweak
and QCD corrections is dominated by the Coulomb-like singularity close
to threshold while turning negative and reaching about $-15\%$ at high
energies.

\section{Summary}

Recently, the full $\Oa$ electroweak corrections have become available
for the Higgs-production processes $\eennh$ and $\eetth$.  In both
cases at least two completely independent calculations have been
performed by different groups, agreeing better than $0.3\%$. The
corrections are sizeable and can reach $\pm10\%$ and will thus be an
important ingredient of precise theoretical predictions for future
$\ee$ colliders.

\end{document}